\documentclass{aa}
\usepackage{graphicx}

\newcommand{\change}{}

\begin{document}

\title{Measurements of magnetic fields over the pulsation cycle in six roAp stars with FORS\,1 at the VLT
\thanks{Based on observations obtained at the European Southern Observatory, Paranal, Chile (ESO 
programmes Nos.~69.D-0210 and 270.D-5023)}}

\author{S. Hubrig \inst{1}\and D.W. Kurtz \inst{2}\and S. Bagnulo 
\inst{1}\and T. Szeifert \inst{1}\and M. Sch\"oller
\inst{1}\and G. Mathys \inst{1}\and W.A. Dziembowski \inst{3}}

\institute{European Southern Observatory, Casilla 19001, Santiago 19, Chile
\and Centre for Astrophysics, University of Central Lancashire, Preston, PR1 2HE, UK
\and Warsaw University Observatory, Al. Ujazdowskie, 4, PL-00-478 Warsawa, Poland
}

\date{Received xx/Accepted yy}

\offprints{S. Hubrig}

\titlerunning{Measurements of magnetic fields over the pulsation cycles}
\authorrunning{Hubrig et al.}

\abstract{With FORS\,1 at the VLT we have tried for the first time to measure the magnetic field variation over 
the pulsation cycle in six roAp stars to begin the study of how the magnetic field and pulsation interact. 
For the star HD\,101065, which has one of the highest photometric pulsation amplitudes of any roAp star, we 
found a signal at the known photometric pulsation frequency at the
3$\sigma$ level in one data set; however this could not be confirmed
by later 
observations. A preliminary simple calculation of the expected magnetic variations 
over the pulsation cycle suggests that they are of the same order as
our current noise levels, leading us to expect that further  
observations with increased S/N have a good chance of achieving an unequivocal detection.


}

\maketitle

\section{Introduction}\label{sec1}
The rapidly oscillating stars (roAp) were the first stars for which solar-like high-overtone p-mode pulsations 
have been definitely detected and are, therefore, prime candidates for asteroseismological studies. The roAp 
stars are cool chemically peculiar stars that pulsate in high-overtone ($n\gg{}l$), low-degree ($l\le3$) 
$p$-modes with periods from about 6 to 15\,min and typical photometric $B$ amplitudes of a few mmag. There are 
32 such stars currently known. Detailed reviews of the roAp stars have been published by Kurtz (\cite{K90}), 
Matthews (\cite{M91}), and Kurtz \& Martinez (\cite{KM00}).

The pulsations of roAp stars are to a large extent governed by their magnetic field (Bigot et al. \cite{B00}; 
Cunha \& Gough \cite{CG00}). An effect of magnetic field on p-mode oscillations has been recently considered
by Bigot \& Dziembowski (\cite{BD02}). 
Their theory of the interaction of rotation, pulsation and the magnetic field suggests a new model, the 
improved oblique pulsator model, that is a significant departure from the standard oblique pulsator model for 
roAp stars. They suggest that the light 
variations are caused by a pulsation mode in which the stellar surface moves in a plane that is 
inclined to both the rotation and the magnetic axes of the star.
The displacement vector describes an ellipse in that plane with the pulsation period, and the whole pattern 
rotates with the rotation of the oblique magnetic field.

From the theoretical considerations, a simple estimate shows that a pulsationally--modulated variation of the
order of $\approx 10^2$\,G may exist in the outer atmospheric layers of roAp stars with kG magnetic fields. 
Assuming that the unperturbed field is nearly force-free and ignoring
the angular derivatives of the displacement vector $\vec\xi$, 
we have in a good approximation for the magnetic field amplitude
$$\delta B_\theta=B_r{\partial\xi_\theta\over\partial r}-B_\theta{\partial\xi_r\over\partial r}$$ and 
$$|\delta B_r|\ll|\delta B_\theta|.$$
The local plane-parallel approximation for  $\vec\xi$ is well justified in our
case of high-frequency low-degree modes in outer layers. 
%{\change which is an excellent approximation
%for low--$l$, high frequency modes, and not only for plan parallel radial pulsations},
%we have in a good approximation for the magnetic field amplitude
%$$\delta B_\theta=B_r{\partial\xi_\theta\over\partial r}-B_\theta{\partial\xi_r\over\partial r}$$ and 
%$$|\delta B_r|\ll|\delta B_\theta|.$$ 
In the same approximation 
we have for the perturbed density
$${\delta\rho\over\rho}=-{\partial\xi_r\over\partial r}.$$
%If the effect of the field on $\vec\xi$ is ignored, then
%$|\xi_\theta|\ll\xi_r$ but this is not justified in the outer
%layers of Ap stars. Using only $|\xi_\theta|\sim\xi_r$, which is
%safe, we get then the following crude estimate: 
%$${\left|\delta B_\theta \over B\right|} \sim\left|{\partial\xi\over\partial
%r}\right|\sim\left|{\delta\rho\over\rho}\right|.$$ 
From these assumptions we get the following crude estimate:
$${\left|\delta B_\theta \over B\right|} \sim\left|{\delta\rho\over\rho}\right|.$$ 
Both, numerical nonadiabatic calculations of oscillations in nonmagnetic 
stars and the quasi-adiabatic approximation suggest that the magnitude of the relative 
intensity amplitude should be of the same order as that of the density.
If we assume the same for magnetic stars which seems reasonable  
then a 10 mmag in light amplitude
translates to 1\%  variations of the field.
This number refers to the subphotosperic layer where
the light amplitude is fixed.
An estimate for the expected field 
variations in the atmosphere may be obtained 
only with radial velocity data. 
For example, from the amplitude rise of
$\delta V\sim$0.8\,km\,s$^{-1}$ in the atmosphere ($h\sim$1000\,km)
we would get 
$$\left|{\partial\xi\over\partial r}\right|\sim{P\delta V\over2h\pi}\sim0.08$$ where $P$ is the 
pulsation period.
{\change Considering that non-adiabatic oscillations in magnetic stars have not been modelled, our 
calculation of the possible pulsationally-modulated variation of the magnetic field is simply a first 
order-of-magnitude estimate of the effect that may be expected.} 
The adopted  radial velocity amplitude rise of $\sim$0.8\,km\,s$^{-1}$ is based on the radial 
velocity measurements of 
\ion{Pr}{iii} line at $\lambda$6160\,\AA{} in
$\gamma$\,Equ (Kochukhov \& Ryabchikova (\cite {KR01a}). This estimate implies 
magnetic field variations at the level of 8\%. 
In general, a large pulsational amplitude is found in doubly ionized rare--earth ions, although a 
large pulsational amplitude in the H$_\alpha$ line, up to 1.46\,km\,s$^{-1}$, has also been reported for the 
roAp star HD\,83368 (Balona \cite{BA02}). The pulsational
velocity amplitudes in the H$_\beta$ and H$_\gamma$ lines in the spectra of this star show, however, much 
lower amplitudes. 
For the star HD\,128898, Kochukhov \& Ryabchikova (\cite {KR01b}) detected modulation of the radial velocity 
amplitude in the \ion{Nd}{iii} line at $\lambda$6145\,\AA{} with a variation between 300 and 500\,m\,s$^{-1}$.
The pulsations in the  radial velocities of the rare--earth lines in this star have been confirmed by 
Balona \& Zima (\cite{BZ02}), but they were only very weakly present in the radial velocities of the
hydrogen lines.
An analysis of individual spectral lines in the roAp star HD\,137949 revealed the pulsational radial velocity 
amplitude ranging from 320\,m\,s$^{-1}$ to the amplitude as low as 7\,m\,s$^{-1}$ (Hatzes et al. \cite {HKM99}).
To summarize, the available radial velocity data imply an estimate for the expected 
magnetic field variations in the atmospheres of roAp stars at the level of 1 to 14\%.

{\change This estimate is given here as an illustration of the fact
that some theoretical arguments suggest that pulsation may induce
magnetic field variations that will be detectable through
observation with currently existing instruments. This represents
the motivation underlying the present study. Admittedly,
ambiguities and unknowns remain in our current understanding
of the physics of pulsation in roAp stars, and possible
alternative theoretical interpretations may lead to somewhat
different estimates of the magnetic field variations to be
expected -- some possibly below the detection threshold.
Accordingly, observations of the latter may provide
a useful discriminant between various models that are currently
possible.}

Despite of importance of magnetic fields for the proper understanding of pulsational properties of roAp stars,
these fields have scarcely been studied until now (Mathys \cite{M03};
Hubrig et al. \cite{H03b}). How this pulsation pattern interacts with the magnetic field 
on the time-scale of the pulsations was entirely untested observationally before our VLT runs in April and 
September 2002. 
{\change The primary purpose of this paper is to provide observational constraints on such magnetic
field variations.}
Because of the short duration of the pulsations (6--15\,min), such observations could not be 
obtained in the past with 4-m class telescopes. The required integration time for measuring the magnetic 
field in a $V=8$ magnitude star with UT1 and FORS1 is only a few tens of seconds. This has allowed us to carry
out an entirely new kind of observation to search for magnetic variability over the pulsation cycle.

Here we present new results of measurements of the mean longitudinal field of six roAp stars obtained from 
low resolution spectropolarimetry with FORS1 at the VLT. 
\section{Observations and data reduction}\label{sec2}

\begin{figure}
\centering
\includegraphics[width=0.45\textwidth]{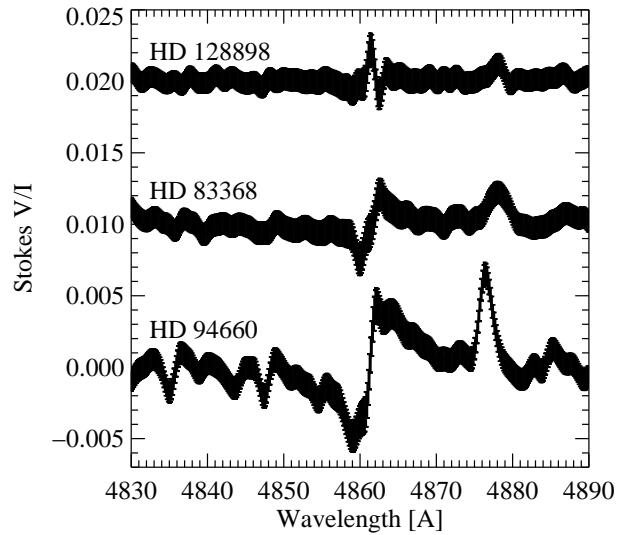}
\caption{
Stokes~{\it V} profiles in the region of the H$_\beta$ line in the two roAp stars HD\,128898 and HD\,83368 and 
in the strongly magnetic Ap star HD\,94660. The spectra of HD\,83368 and HD\,128898 are shifted in the 
vertical direction by 0.01 and 0.02 respectively. The thickness of the plotted lines corresponds to the 
uncertainty of the measurement of polarization determined from photon noise. A strong polarization signal 
close to the wavelength $\lambda \approx 4876$\,\AA{} is probably due to the presence of the blend of 
strong lines belonging to \ion{Nd}{ii} and \ion{Cr}{ii}.}
\label{fig:feature}
\end{figure}

Time series of spectra of bright roAp stars were obtained with FORS\,1
(FOcal Reducer low dispersion Spectrograph) mounted on the 8-m Melipal
(UT3) telescope of the VLT in April 2002 (HD\,83368, HD\,101065, HD\,128898 and HD\,137949) 
and in September 2002 (HD\,201601 and HD\,217522).
The FORS\,1 multi-mode instrument is equipped with polarization analyzing optics comprising super-achromatic 
half-wave and
quarter-wave phase retarder plates, and a Wollaston prism with a beam divergence of 22$\arcsec$ in standard 
resolution mode (Appenzeller et al. \cite{AF98}). During our April run we used the GRISM\,600B to cover all H 
Balmer lines from H$_\beta$ to the Balmer jump, and during the
September run we observed with GRISM\,600R in the region which 
includes H$_\alpha$ and H$_\beta$, from 4770 to 6900\,\AA. Both grisms have 
600 grooves mm$^{-1}$; with the narrowest available slit width of 0$\farcs$4 they give a spectral resolving power of R$\sim$2000 and R$\sim$2900, respectively. 
Wavelength calibrations were taken during day time for the two different retarder waveplate setups 
($\alpha$=$+$45$^\circ$ and $-$45$^\circ$) which are the same as those used for the observations. Wavelength 
calibration was performed by associating with each individual science spectrum the calibration frame obtained 
with the similar orientation of the retarder waveplate. Ordinary and extraordinary beams were independently 
calibrated with the corresponding beams of the reference spectrum. As has been previously shown by 
Landstreet (\cite{L82}), in the weak field regime, the mean longitudinal field can be derived from the 
difference between the circular polarizations observed in the red and blue wings of the hydrogen line profiles
using the formula 
\begin{equation}
\frac{V}{I} = -\frac{g_{\rm eff} e \lambda^2}{4\pi{}m_ec^2}\ \frac{1}{I}\ \frac{{\rm d}I}{{\rm d}\lambda} \left<{\cal B}_z\right>,
%\label{eq:one}
\end{equation}
where $V$ is the Stokes parameter which measures the circular polarization, $I$ is the intensity in the 
unpolarized spectrum, $g_{\rm eff}$ is the effective Land\'e factor, $e$ is the electron charge, $\lambda$ is 
the wavelength, 
%expressed in \AA{}, 
$m_e$ the electron mass, $c$ the speed of light, 
% the following is trivial and useless:
% ${{\rm d}I/{\rm d}\lambda}$ is the derivative of Stokes $I$, 
and $\left<{\cal B}_z\right>$ is the mean longitudinal field.
% expressed in Gauss. 
To minimize the cross-talk effect we executed the sequence $+45-45$, $+45-45$, $+45-45$ etc. and 
calculated the values $V/I$ using: 
\begin{equation}
\frac{V}{I} = \frac{1}{2} \left\{ \left( \frac{f^{\rm o} - f^{\rm e}}{f^{\rm o} + f^{\rm e}} \right)_{\alpha=-45^{\circ}} - \left( \frac{f^{\rm o} - f^{\rm e}}{f^{\rm o} + f^{\rm e}} \right)_{\alpha=+45^{\circ}} \right\},
%\label{eqn:two} 
\end{equation}
where $\alpha$ gives the position angle of the retarder waveplate and $f^{\rm o}$ and $f^{\rm e}$ are ordinary
and extraordinary beams, respectively. 
%Stokes $I$ values have been obtained from the sum of the ordinary and extraordinary beams. 
%To derive $\left<{\cal B}_z \right>$ a least-squares technique has been used to minimize the expression
%\begin{displaymath}
%\chi^2 = \sum_i \frac{(y_i - \left< {\cal B}_z \right> x_i - b)^2}{\sigma_i^2}
%\end{displaymath}
%where, for each spectral point $i$, $y_i = (V/I)_i$,
%$x_i = -[g_{\rm eff}\,e\,\lambda_i^2/(4\pi\,m_e\,c^2)\,(1/I\ \times\ {\rm d}I/{\rm d}\lambda)_i$
%and $b$ is a constant term that, 
% GM: I do not understand what follows. Presumably, other readers will
% not understand either. 
%assuming that Eq.~(1) is correct, approximates the fraction of instrumental polarization not removed after th
%e application of Eq.~(2) to the observations. For each spectral point $i$, the derivative of Stokes $I$ with r
%espect to the wavelength was evaluated as 
%\begin{displaymath}
%\left( \frac{{\rm d}I}{{\rm d}\lambda} \right)_{\lambda=\lambda_i} = \frac{N_{i+1}-N_{i-1}}{\lambda_{i+1}-\lambd%a_{i-1}},
%\end{displaymath}
%where $N_i$ is the photon count at wavelength $\lambda_i$. 
In our calculations we assumed a Land\'e factor $g_{\rm eff}$=1 for hydrogen lines and $g_{\rm eff}$=1.25 for metal lines. Table~\ref{tab:balmer} lists the wavelength ranges corresponding to the hydrogen Balmer lines for 
which the Land\'e factor has been set to 1. 
Furthermore, in our reduction procedure the spectral regions containing telluric lines have been excluded in 
the measurements of magnetic field. 
\begin{table}
\caption{
\label{tab:balmer}
Wavelength ranges around the hydrogen Balmer lines for which the Land\'e
factor has been set to 1. In all other wavelength ranges the Land\'e factor $g_{\rm eff}$=1.25 has 
been adopted. 
}
\begin{center}
\begin{tabular}{cc}
Line(s) & Wavelength range [\AA{}] \\
\hline
H$_{16}$-H$_{10}$ & 3701 -- 3801 \\
H$_9$ & 3820.5 -- 3852.5 \\
H$_8$ & 3870.2 -- 3910.2 \\
H$_\epsilon$ & 3941.2 -- 4001.2 \\
H$_\delta$ & 4082.9 -- 4122.9 \\
H$_\gamma$ & 4311.7 -- 4371.7 \\
H$_\beta$ & 4812.7 -- 4912.7 \\
H$_\alpha$ & 6512 -- 6612 \\
\end{tabular}
\end{center}
\end{table}
More details of the observing technique are given by Bagnulo et al. (\cite{B02}) and Hubrig et al. 
(\cite{H03a}). The errors of the measurements of the polarization have been determined from photon counting 
statistics and have been converted to errors of field measurements. On each night an additional star with a 
well-defined strong longitudinal field was selected to check that the
instrument was functioning properly. The  
star HD\,94660, which was observed in April, has a longitudinal
magnetic field that varies about a mean value of $\sim-1900$~G with a low amplitude ($\sim160$~G
peak-to-peak) over a period of 2800~d (Mathys et al. \cite{MMW03}).
The result of our measurement 
using the whole spectral region from 3500 to 5800\,\AA, $\left<{\cal B}_z\right>$=$-2056\pm$19\,G, is fully 
consistent with the value of the longitudinal field at the considered
rotation phase that is expected from the variation curve defined by
Mathys et al. (\cite{MMW03}). The measurements obtained from  
the individual Balmer lines show slightly different results. For instance, using for the measurements 
exclusively the H$_\beta$ line we obtain $\left< {\cal B}_z\right>_{H_{\beta}} = -2442\pm$73\,G. The reasons 
for different values of $\left<{\cal B}_z\right>$ derived from Balmer lines in comparison to the values 
obtained using the whole region are discussed in more detail by Bagnulo et al.\ (\cite{B02}). At this point we
would like to mention that for all stars of our sample the magnetic field measured in the wings of H$_\beta$ 
% GM: you have inconsistent notations for the Balmer lines throughout
% the paper: e.g., sometimes you use H$\beta$ (which I believe is
% correct), and in other places you use H$_\beta$ (which I think is
% not right). I have just corrected the previous instance, but you
% should homogeneise this throughout the whole paper (starting with
% next line...).  
is systematically larger, up to 500\,G, than the field derived from the H$_\gamma$, H$_\delta$ and H$_\epsilon$ 
lines. The inspection of the behaviour of the polarized spectrum in the wings of the H$_\beta$ line reveals a 
very strong signal at the wavelength $\lambda \approx 4876$\,\AA{} where the blend of strong lines 
% GM: according to the A&A instructions, the proper way to refer to an
% ion is using their macro \ion (as shown below). I changed this in
% several instances, but I may have overlooked some. You should check.
belonging to \ion{Nd}{ii} and \ion{Cr}{ii} is located (Fig.~\ref{fig:feature}). This feature probably accounts 
for the excess of magnetic field measured in the region of H$_\beta$. However, high resolution 
spectropolarimetric observations of these stars are needed to verify the origin of such a strong polarization
signal.

% GM: for me this is not obvious at all. The only convincing way to
% argue about this would be to compare a measurement including the
% blend with one that does not include it. 

The star HD\,187474, which has a rotation period of 6.4~yr and a longitudinal magnetic field ranging from 
$\approx -2$\,kG to 2\,kG, was observed during the night of 14 September 2002 at the rotation phase 0.46. The 
derived value of the magnetic field, $\left<{\cal B}_z\right> = -1849\pm$47\,G, fits very well to the 
observations at the same phase presented by Mathys et al. (\cite{MHL97}).

The individual measurements of the longitudinal magnetic field for each star are presented in 
Tables~\ref{tab:hd83368}~to~\ref{tab:hd217522}. After finishing the time series for each star in our sample of 
four stars in the observing night in April we decided to use the GRISM 600R for a few exposures to check the 
consistency of the magnetic field measurements in the region including the H$_\alpha$ line with that which 
contains all other Balmer lines. No systematic difference has been found between the measurements in two 
different spectral regions.

% GM: I recommend that you modify the preamble of Tables 3 to 8 as I
% have done below for Table 2, for better lisibility. 
\begin{table}
\caption{
\label{tab:hd83368}
Mean longitudinal magnetic field for the time series of HD\,83368.
}
\begin{center}
\begin{tabular}{crr|crr}
\hline
MJD &\multicolumn{1}{c}{$\left<{\cal B}_z\right>$}&\multicolumn{1}{c}{$\sigma{}_z$} 
&MJD&\multicolumn{1}{c}{$\left<{\cal B}_z\right>$} &\multicolumn{1}{c}{$\sigma{}_z$}\\
&\multicolumn{1}{c}{[G]}&\multicolumn{1}{c}{[G]} & &\multicolumn{1}{c}{[G]}&\multicolumn{1}{c}{[G]} \\
\hline
52382.977731 & $-$868 & 60 & 	52383.039369 & $-$786 & 54\\
52382.979433 & $-$861 & 60 & 	52383.041174 & $-$854 & 57\\
52382.981137 & $-$805 & 60 & 	52383.042984 & $-$818 & 60\\
52382.982841 & $-$608 & 60 & 	52383.044806 & $-$829 & 60\\
52382.984553 & $-$1057& 63 & 	52383.046728 & $-$879 & 47\\
52382.986283 & $-$788 & 57 & 	52383.048638 & $-$720 & 51\\
52382.988050 & $-$777 & 54 & 	52383.050505 & $-$830 & 54\\
52382.989816 & $-$783 & 60 & 	52383.052323 & $-$990 & 57\\
52382.991590 & $-$879 & 60 & 	52383.054130 & $-$718 & 60\\
52382.993386 & $-$829 & 54 & 	52383.055971 & $-$681 & 54\\
52382.995213 & $-$839 & 51 & 	52383.057846 & $-$920 & 51\\
52382.997032 & $-$845 & 54 & 	52383.059721 & $-$833 & 54\\
52382.998804 & $-$892 & 57 & 	52383.061596 & $-$932 & 54\\
52383.000575 & $-$771 & 57 & 	52383.063473 & $-$764 & 51\\
52383.002350 & $-$782 & 57 & 	52383.065349 & $-$821 & 51\\
52383.006425 & $-$1181 & 54 & 	52383.069270 & $-$805 & 54\\
52383.008209 & $-$905 & 57 & 	52383.071105 & $-$848 & 51\\
52383.009994 & $-$830 & 54 & 	52383.072939 & $-$910 & 51\\
52383.011779 & $-$575 & 57 & 	52383.074751 & $-$790 & 57\\
52383.013568 & $-$897 & 54 & 	52383.076566 & $-$965 & 51\\
52383.015354 & $-$1115 & 54 & 	52383.078386 & $-$864 & 51\\
52383.017147 & $-$767 & 54 & 	52383.080183 & $-$1011 & 57\\
52383.018943 & $-$940 & 57 & 	52383.081960 & $-$896 & 57\\
52383.020728 & $-$820 & 60 & 	52383.083742 & $-$817 & 57\\
52383.022515 & $-$957 & 60 & 	52383.085520 & $-$920 & 57\\
52383.024320 & $-$834 & 60 & 	52383.087300 & $-$801 & 57\\
52383.026218 & $-$955 & 51 & 	52383.089080 & $-$868 & 60\\
52383.028119 & $-$687 & 54 & 	52383.090877 & $-$812 & 60\\
52383.029930 & $-$817 & 60 & 	52383.092768 & $-$863 & 57\\
52383.031730 & $-$950 & 54 & 	52383.094707 & $-$635 & 51\\
\hline
\end{tabular}
\end{center}
\end{table}

\begin{table}
\caption{
\label{tab:hd101065a}
Mean longitudinal magnetic field for the time series of HD\,101065
obtained in April 2002.
}
\begin{center}
\begin{tabular}{crr|crr}
\hline
MJD &\multicolumn{1}{c}{$\left<{\cal B}_z\right>$}&\multicolumn{1}{c}{$\sigma{}_z$} 
&MJD&\multicolumn{1}{c}{$\left<{\cal B}_z\right>$} &\multicolumn{1}{c}{$\sigma{}_z$}\\
&\multicolumn{1}{c}{[G]}&\multicolumn{1}{c}{[G]} & &\multicolumn{1}{c}{[G]}&\multicolumn{1}{c}{[G]} \\
\hline
52383.148501 & $-$1061 & 51 &	52383.202286 & $-$914 & 47 \\
52383.151370 & $-$1072 & 51 &	52383.205434 & $-$1032 & 51 \\
52383.154296 & $-$1075 & 51 &	52383.208556 & $-$949 & 51 \\
52383.157568 & $-$968 & 44 &	52383.211683 & $-$1023 & 51 \\
52383.160875 & $-$1105 & 44 &	52383.214809 & $-$1015 & 54 \\
52383.164007 & $-$1128 & 51 &	52383.217932 & $-$1068 & 54 \\
52383.167110 & $-$926 & 51 &	52383.221058 & $-$1021 & 54 \\
52383.170237 & $-$1075 & 51 &	52383.224218 & $-$1043 & 57 \\
52383.173541 & $-$1025 & 51 &	52383.227554 & $-$1075 & 51 \\
52383.176883 & $-$1072 & 54 &	52383.230917 & $-$1119 & 51 \\
52383.180222 & $-$1060 & 54 &	52383.234282 & $-$1019 & 54 \\
52383.183554 & $-$999 & 51 &	52383.237652 & $-$1085 & 54 \\
52383.186888 & $-$1105 & 51 &	52383.241020 & $-$999 & 57 \\
52383.190224 & $-$1052 & 47 &	52383.244393 & $-$1017 & 60 \\
52383.193564 & $-$1063 & 47 &	52383.247762 & $-$1062 & 60 \\
\hline
\end{tabular}
\end{center}
\end{table}

\begin{table}
\caption{
\label{tab:hd128898}
Mean longitudinal magnetic field for the time series of HD\,128898.
}
\begin{center}
\begin{tabular}{crr|crr}
\hline
MJD &\multicolumn{1}{c}{$\left<{\cal B}_z\right>$}&\multicolumn{1}{c}{$\sigma{}_z$} 
&MJD&\multicolumn{1}{c}{$\left<{\cal B}_z\right>$} &\multicolumn{1}{c}{$\sigma{}_z$}\\
&\multicolumn{1}{c}{[G]}&\multicolumn{1}{c}{[G]} & &\multicolumn{1}{c}{[G]}&\multicolumn{1}{c}{[G]} \\
\hline
52383.278303 & $-$5 & 60 &	52383.301002 & $-$306 & 63 \\
52383.279870 & $-$361 & 60 &	52383.302584 & $-$1 & 66 \\
52383.281422 & $-$63 & 63 &	52383.304163 & $-$195 & 70 \\
52383.282986 & $-$368 & 63 &	52383.305740 & $-$385 & 70 \\
52383.284544 & $-$291 & 63 &	52383.307315 & $-$287 & 70 \\
52383.288427 & $-$163 & 60 &	52383.308895 & $-$342 & 63 \\
52383.289992 & $-$256 & 63 &	52383.310474 & $-$233 & 70 \\
52383.291561 & $-$478 & 63 &	52383.314201 & $-$230 & 57 \\
52383.293135 & $-$472 & 63 &	52383.315821 & $-$148 & 54 \\
52383.294707 & $-$298 & 60 &	52383.317441 & $-$251 & 57 \\
52383.296279 & $-$321 & 63 &	52383.319067 & $-$23 & 60 \\
52383.297857 & $-$40 & 63 &	52383.320686 & $-$310 & 57 \\
52383.299426 & $-$118 & 63 &	52383.322304 & $-$271 & 57 \\
\hline
\end{tabular}
\end{center}
\end{table}

\begin{table}
\caption{
\label{tab:hd137949}
Mean longitudinal magnetic field for the time series of HD\,137949.
}
\begin{center}
\begin{tabular}{crr|crr}
\hline
MJD &\multicolumn{1}{c}{$\left<{\cal B}_z\right>$}&\multicolumn{1}{c}{$\sigma{}_z$} 
&MJD&\multicolumn{1}{c}{$\left<{\cal B}_z\right>$} &\multicolumn{1}{c}{$\sigma{}_z$}\\
&\multicolumn{1}{c}{[G]}&\multicolumn{1}{c}{[G]} & &\multicolumn{1}{c}{[G]}&\multicolumn{1}{c}{[G]} \\
\hline
52383.341268 & 2111 & 47 &	52383.371795 & 2193 & 51 \\
52383.343093 & 2161 & 47 &	52383.373714 & 2003 & 54 \\
52383.344925 & 2204 & 51 &	52383.375634 & 2092 & 51 \\
52383.346754 & 2078 & 51 &	52383.377559 & 2279 & 54 \\
52383.348591 & 2154 & 54 &	52383.379490 & 2107 & 54 \\
52383.350428 & 2132 & 51 &	52383.381451 & 2105 & 54 \\
52383.352264 & 2154 & 57 &	52383.383430 & 2108 & 51 \\
52383.354154 & 2121 & 51 &	52383.385404 & 2140 & 57 \\
52383.356050 & 2246 & 51 &	52383.387515 & 2190 & 54 \\
52383.357898 & 2154 & 54 &	52383.389743 & 2164 & 47 \\
52383.359796 & 2159 & 47 &	52383.391962 & 2139 & 47 \\
52383.361710 & 2223 & 47 &	52383.394179 & 2072 & 47 \\
52383.363619 & 2162 & 47 &	52383.396398 & 2168 & 47 \\
52383.365523 & 2138 & 47 &	52383.398609 & 2113 & 47 \\
52383.367434 & 2198 & 51 &	52383.400823 & 2132 & 47 \\
\hline
\end{tabular}
\end{center}
\end{table}

\begin{table}
\caption{
\label{tab:hd201601}
Mean longitudinal magnetic field for the time series of HD\,201601.
}
\begin{center}
\begin{tabular}{crr|crr}
\hline
MJD &\multicolumn{1}{c}{$\left<{\cal B}_z\right>$}&\multicolumn{1}{c}{$\sigma{}_z$} 
&MJD&\multicolumn{1}{c}{$\left<{\cal B}_z\right>$} &\multicolumn{1}{c}{$\sigma{}_z$}\\
&\multicolumn{1}{c}{[G]}&\multicolumn{1}{c}{[G]} & &\multicolumn{1}{c}{[G]}&\multicolumn{1}{c}{[G]} \\
\hline
52530.991626 & $-$1081 & 60 & 52531.044233 & $-$1086 & 57 \\
52530.995482 & $-$1135 & 54 & 52531.047133 & $-$915 & 57 \\
52531.017022 & $-$842 & 47 & 52531.051108 & $-$1112 & 54 \\
52531.026988 & $-$1065 & 57 & 52531.054008 & $-$1106 & 54 \\
52531.029854 & $-$739 & 57 & 52531.056909 & $-$930 & 51 \\
52531.032729 & $-$1164 & 60 & 52531.059824 & $-$1084 & 51 \\
52531.035599 & $-$1157 & 57 & 52531.084795 & $-$1286 & 79 \\
52531.038471 & $-$1273 & 51 & 52531.093915 & $-$777 & 47 \\
52531.041351 & $-$1163 & 54 & 52531.096943 & $-$1384 & 51 \\
\hline
\end{tabular}
\end{center}
\end{table}

\begin{table}
\caption{
\label{tab:hd217522}
Mean longitudinal magnetic field for the time series of HD\,217522.
}
\begin{center}
\begin{tabular}{crr|crr}
\hline
MJD &\multicolumn{1}{c}{$\left<{\cal B}_z\right>$}&\multicolumn{1}{c}{$\sigma{}_z$} 
&MJD&\multicolumn{1}{c}{$\left<{\cal B}_z\right>$} &\multicolumn{1}{c}{$\sigma{}_z$}\\
&\multicolumn{1}{c}{[G]}&\multicolumn{1}{c}{[G]} & &\multicolumn{1}{c}{[G]}&\multicolumn{1}{c}{[G]} \\
\hline
52531.108661 & $-$701 & 79 &	52531.222581 & $-$1123 & 82 \\
52531.112774 & $-$822 & 79 &	52531.224908 & $-$937 & 107 \\
52531.117013 & $-$862 & 70 &	52531.241321 & $-$468 & 76 \\
52531.121656 & $-$765 & 70 &	52531.243544 & $-$695 & 76 \\
52531.123913 & $-$789 & 70 &	52531.245767 & $-$588 & 79 \\
52531.126157 & $-$808 & 70 &	52531.247992 & $-$696 & 79 \\
52531.128421 & $-$574 & 70 &	52531.250215 & $-$672 & 85 \\
52531.130694 & $-$709 & 73 &	52531.252442 & $-$717 & 85 \\
52531.132988 & $-$640 & 70 &	52531.254668 & $-$425 & 79 \\
52531.135257 & $-$689 & 73 &	52531.256897 & $-$734 & 76 \\
52531.137511 & $-$895 & 88 &	52531.259126 & $-$676 & 76 \\
52531.140799 & $-$696 & 79 &	52531.261359 & $-$651 & 79 \\
52531.143058 & $-$655 & 70 &	52531.263593 & $-$621 & 79 \\
52531.145320 & $-$728 & 73 &	52531.265828 & $-$705 & 79 \\
52531.147573 & $-$747 & 92 &	52531.268061 & $-$655 & 82 \\
52531.149827 & $-$763 & 88 &	52531.270295 & $-$629 & 76 \\
52531.152081 & $-$682 & 79 &	52531.272531 & $-$586 & 82 \\
52531.154333 & $-$764 & 73 &	52531.274774 & $-$700 & 82 \\
52531.156587 & $-$629 & 79 &	52531.277904 & $-$718 & 79 \\
52531.158861 & $-$708 & 73 &	52531.280145 & $-$661 & 79 \\
52531.161141 & $-$699 & 73 &	52531.282384 & $-$587 & 82 \\
52531.163417 & $-$512 & 70 &	52531.284626 & $-$758 & 85 \\
52531.165710 & $-$579 & 70 &	52531.286870 & $-$626 & 82 \\
52531.167976 & $-$671 & 76 &	52531.289116 & $-$784 & 82 \\
52531.170258 & $-$553 & 76 &	52531.291365 & $-$698 & 82 \\
52531.172548 & $-$596 & 76 &	52531.293614 & $-$563 & 85 \\
52531.174977 & $-$746 & 73 &	52531.295863 & $-$638 & 85 \\
52531.178151 & $-$812 & 70 &	52531.298115 & $-$756 & 85 \\
52531.180426 & $-$742 & 70 &	52531.300366 & $-$533 & 85 \\
52531.182705 & $-$727 & 70 &	52531.302620 & $-$662 & 85 \\
52531.184983 & $-$712 & 70 &	52531.304873 & $-$759 & 85 \\
52531.187294 & $-$702 & 63 &	52531.307128 & $-$656 & 92 \\
52531.189575 & $-$661 & 66 &	52531.309384 & $-$535 & 88 \\
52531.191868 & $-$795 & 70 &	52531.311642 & $-$547 & 88 \\
52531.194151 & $-$686 & 70 &	52531.314797 & $-$407 & 85 \\
52531.196442 & $-$567 & 63 &	52531.317059 & $-$595 & 88 \\
52531.198730 & $-$695 & 73 &	52531.319321 & $-$945 & 98 \\
52531.201013 & $-$744 & 73 &	52531.321589 & $-$632 & 92 \\
52531.203297 & $-$655 & 70 &	52531.323855 & $-$728 & 98 \\
52531.205589 & $-$681 & 66 &	52531.332912 & $-$735 & 111 \\
52531.207877 & $-$691 & 66 &	52531.335189 & $-$733 & 98 \\
52531.210168 & $-$694 & 66 &	52531.337471 & $-$767 & 98 \\
52531.212455 & $-$768 & 70 &	52531.339747 & $-$827 & 101 \\
52531.215691 & $-$758 & 70 &	52531.342024 & $-$556 & 98 \\
52531.217987 & $-$592 & 70 &	52531.344306 & $-$614 & 98 \\
52531.220277 & $-$424 & 73 &	& & \\
\hline
\end{tabular}
\end{center}
\end{table}

\section{Results for individual stars}\label{sec3}

\begin{figure}
\centering
\includegraphics[width=0.45\textwidth]{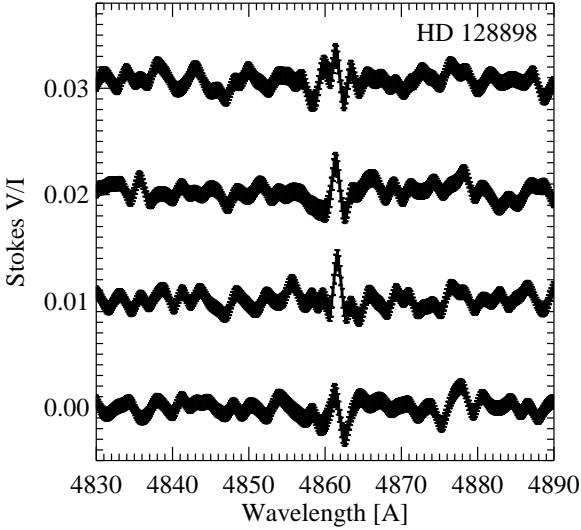}
\caption{
Pulsational behaviour of the Stokes~{\it V} profiles in the region of the H$_\beta$ line over four consecutive 
exposures in the star HD\,128898 with the smallest measured longitudinal magnetic field. As in Fig.~1, the 
thickness of the plotted lines corresponds to the uncertainty of the measurements of polarization determined 
from photon noise.}
\label{fig:HD128898}
\end{figure}

\begin{figure}
\centering
\includegraphics[width=0.45\textwidth]{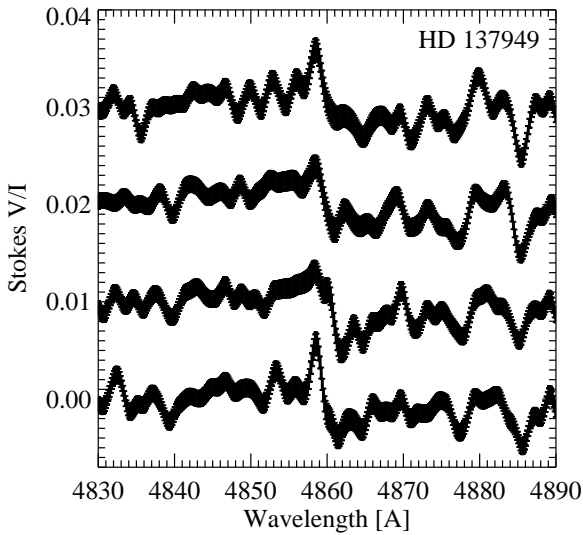}
\caption{
Pulsational behaviour of the Stokes~{\it V} profiles in the region of the H$_\beta$ line over four consecutive 
exposures in the star HD\,137949 with the largest measured longitudinal magnetic field. As in Fig.~1, the 
thickness of the plotted lines corresponds to the uncertainty of the measurements of polarization determined 
from photon noise.}
\label{fig:HD137949}
\end{figure}

For each roAp star we took a continuous series of sets of 2 exposures (with the retarder waveplate oriented at 
different angles), distributed along many consecutive pulsation cycles. As the observed spectral range includes 
all Balmer lines from H$_\beta$ bluewards, the analysis of their Stokes~{\it V} profiles observed in each 
exposure 
permits us to obtain an estimate of the mean longitudinal field with an accuracy better than 100\,G with a few 
seconds of exposure time for HD\,128898, which is the brightest star in our sample, and about 60\,s for the 
faintest star HD\,101065. Among the roAp stars, the selected targets have pulsation periods from 6.8 to 
13.9\,min. Since the exposure time is only a fraction of the pulsation period for all considered stars, by 
Fourier analysing our results over many cycles of pulsation we have been able to examine whether there exist 
variations of the order of a few 100\,G over the pulsation period. Pulsational variability of Stokes~{\it V} 
during four consecutive exposures are presented in
Fig.~\ref{fig:HD128898} for the star HD\,128898, which has the 
smallest measured longitudinal field, and in Fig.~\ref{fig:HD137949}
for the star HD\,137949, which has the largest 
measured longitudinal field.

Given the low resolution of FORS\,1, metallic lines in the spectra of roAp stars appear mostly as unresolved 
blends. Using observations of the strongly magnetic star HD\,94660, Bagnulo et al. (\cite{B02}) studied the 
impact of metallic line blends on the determination of the longitudinal field in comparison to measurements 
of the magnetic field restricted exclusively to the wavelength region containing hydrogen Balmer lines. 
They showed that the determination of the longitudinal field using the full wavelength range, including all 
metallic lines, is reasonably consistent with that obtained within wavelength windows around individual Balmer 
lines.
The roAp stars in our sample are much cooler than HD\,94660 which has the spectral type A0, and their spectra 
are incomparably line rich. 
%with abundances of rare-earth elements enhanced by 3--4\,dex. 
% GM: The above is not correct, or at least it is misleading. Rare
% earth overabundances are as extreme in a hot Ap star such as HD
% 94660 as they are in cooler Ap stars. The cleaner character of the
% spectrum of the former just comes from the fact that the number of
% strong lines of the overabundant elements drops dramatically with
% increasing temperature. Some rewriting is needed here to avoid
% confusion. 
In addition, spectral 
variability throughout the pulsation cycle has been found in some roAp stars. For example, Kochukhov \& 
Ryabchikova (\cite{KR01a}) have shown that in $\gamma$\,Equ the spectral lines of \ion{Pr}{iii} and 
\ion{Nd}{iii} show 
significant radial velocity variations with the pulsation period, while other lines in the spectrum show none. 
Balona (\cite{BA02}) and Balona \& Zima (\cite{BZ02}) get similar results for HD\,83368 and HD\,24712, 
respectively. Interestingly, they have good evidence that the amplitude of the pulsation is higher in 
H$_\alpha$ than in H$_\beta$ and H$_\gamma$ in those stars. Knudsen (\cite {K00}) detected the pulsation modes 
in equivalent width variations of H$_\alpha$ in the star HD\,24712. Bisector measurements of the H$_\alpha$ 
line of the roAp stars HD\,128898 (Baldry et al. \cite{BV99}) and HD\,83368 (Baldry \& Bedding \cite{BB00}) 
have provided pulsational depth information for the first time. Finally, Kurtz, Elkin \& Mathys (\cite{KEM03})
have resolved the magneto-acoustic boundary layer as a function of atmospheric depth in HD\,166473 using high 
time resolution VLT UVES spectra. They obtain radial velocity uncertainties of only 2\,m\,s$^{-1}$, and also 
find no variation in the velocities of Fe lines with velocities up to 80\,m\,s$^{-1}$ in lines of Nd and Pr. The 
derived amplitudes and phases of the radial velocity variations as a
function of depth in roAp stars allow detailed 
atmospheric constraints to be derived.

Because of the suggestion by Balona (\cite{BA02}) that the pulsation amplitude is lower for H$_\beta$ and 
H$_\gamma$ than for H$\alpha$, and because nothing is known about the pulsational behaviour of other Balmer 
lines, Fourier 
analysis has been done using different sets of magnetic field determinations including those in the full 
wavelength range and the measurements within the wavelength windows around the individual Balmer lines 
H$_\alpha$, H$_\beta$, H$_\gamma$ and H$_\delta$. Shortwards from H$_\delta$ the measurements of the magnetic 
field using hydrogen lines show large uncertainties, up to 1000\,G, due to a lower photon count rate and lower 
Zeeman splitting in the blue region.

In the following we present the results of Fourier analyses of magnetic field measurements over several 
pulsation cycles for each star in the sample. Least squares fits have been applied to determine the amplitude 
spectrum of each star. 

\begin{figure}
\centering
\includegraphics[width=0.45\textwidth]{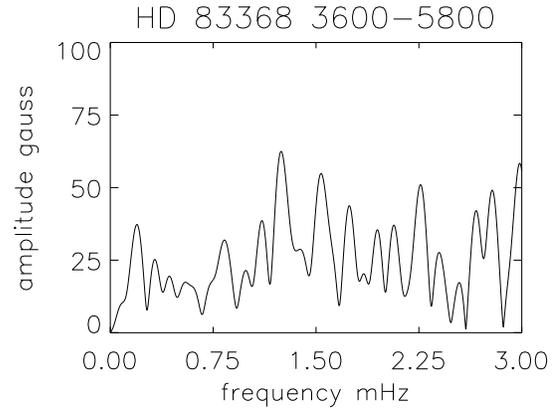}
\caption
{An amplitude spectrum of the magnetic field measurements of HD\,83368 over the entire wavelength 
range studied. This is typical of the amplitude spectra for the stars in the sample. The highest peak is 
at $\nu = 1.246$\, mHz ($P = 13.38$\,min) which is not at the known pulsation frequency of 1.428\,mHz 
($P = 11.67$\,min), hence is probably a noise peak.} 
\label{fig:hd83368}
\end{figure}

{\it \object{HD\,83368}}
(HR\,3831) is singly periodic with a pulsation period of 11.67\,min (Kurtz \cite{K82}). We have 
obtained 60 series of magnetic field measurements with exposure times of 10--18\,s over 14 consecutive 
pulsation cycles. Taking into account the overheads one magnetic field measurement took 
$\null\approx 2.5$\,min. In Fig.~\ref{fig:hd83368} we show the amplitude spectrum of HD\,83368 obtained from the 
measurements in the whole spectral region from 3600 to 5800\,\AA{}. The highest peak at 1.246\,mHz for the 
measurements in the whole spectral region is not at the known photometric pulsation frequency (1.428\,mHz). 
None of the amplitude spectra for the measurements of Balmer lines shows a signal at the known photometric 
pulsation frequency either. It seems highly unlikely that a pulsation frequency
previously undetected in photometric observations, may actually be
present in HD\,83368, which is one of the best studied roAp stars. 
The standard deviation of the measurements is 108\,G -- significantly higher 
than the formal errors given in Table~\ref{tab:hd83368}. This suggests either that there is a signal buried 
in the noise, or that the formal error estimates are too low. 

\begin{figure}
\begin{center}
\resizebox{0.45\textwidth}{!}{\includegraphics{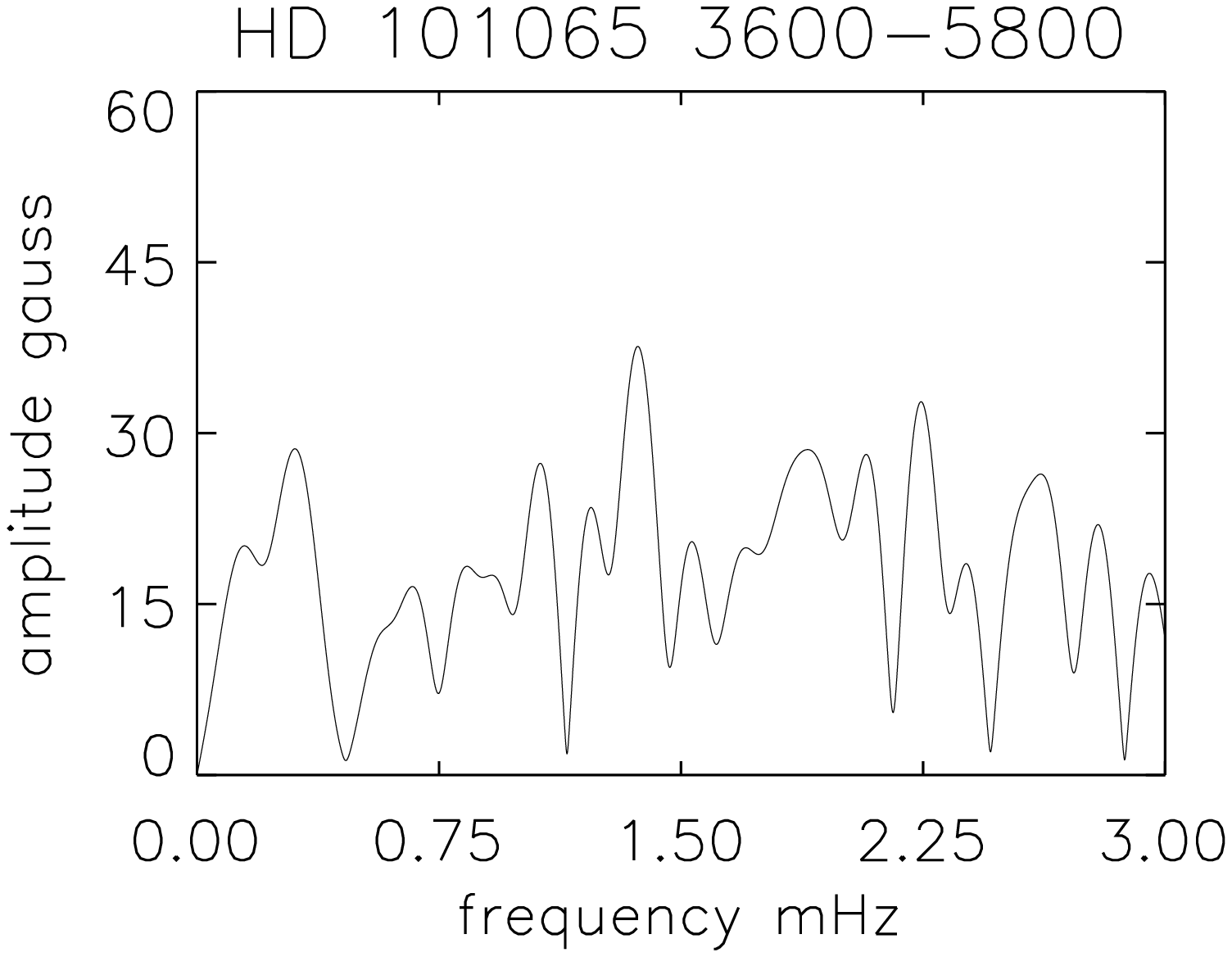}}
\resizebox{0.45\textwidth}{!}{\includegraphics{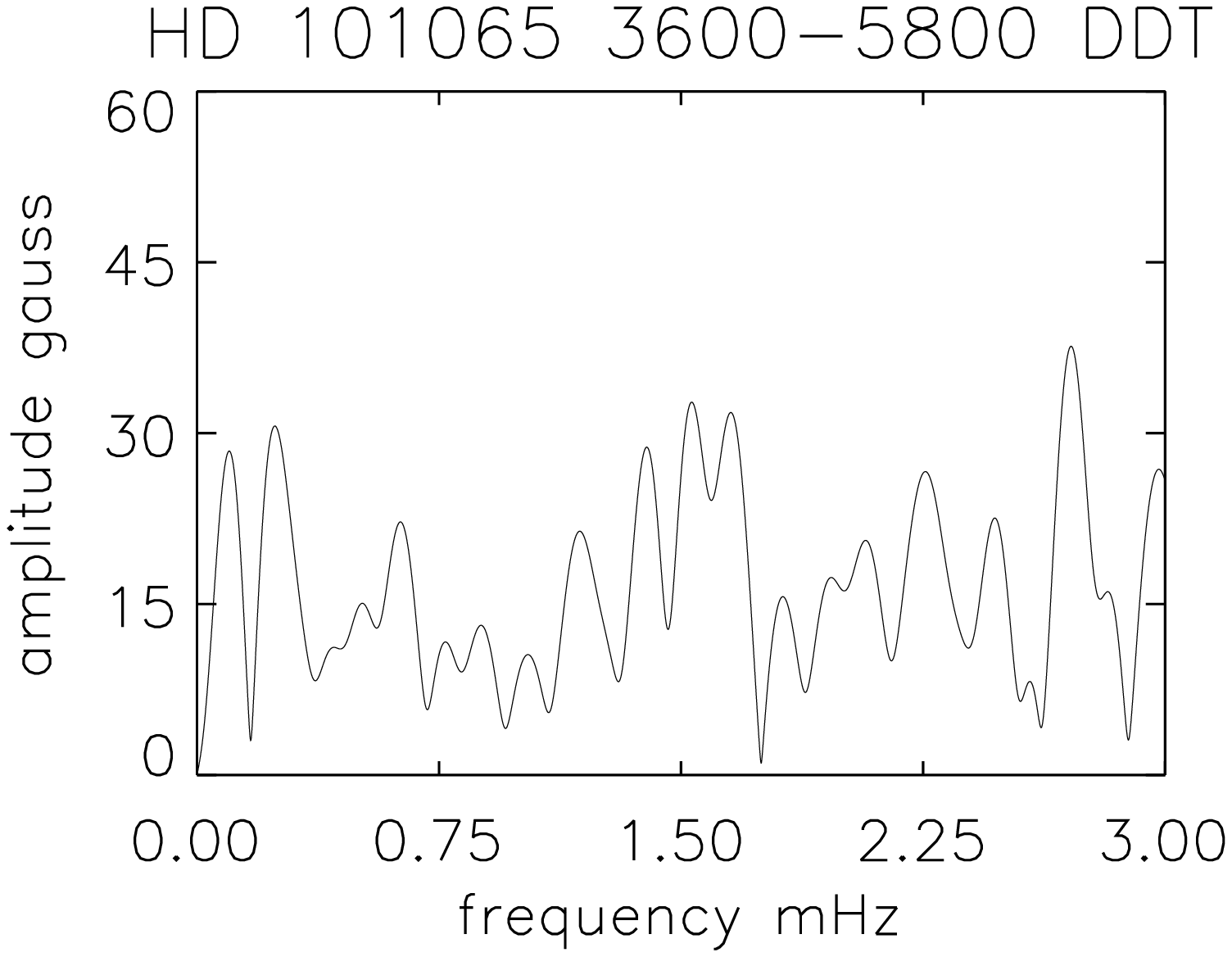}}
\end{center}
\caption{
Amplitude spectra of the magnetic field measurements of HD\,101065. The top panel shows the 
amplitude spectrum for the first 2.4-hr run with the highest peak at the known photometric pulsation 
frequency, 1.365\,mHz, with an amplitude of $39 \pm 12$\,G. The lower panel shows the amplitude spectrum of 
a 4-hr run under Director's Discretionary Time that fails to confirm the highest peak in the upper panel.}
\label{fig:hd101065}
\end{figure}

{\it \object{HD\,101065}}
(Przybylski's star) pulsates with one principal mode with a frequency of $\nu = 1.373$\,mHz 
($P =12.14$\,min) and with two other much lower amplitude frequencies (Martinez \& Kurtz \cite{MK90}). This 
star has the highest photometric pulsation amplitude ($\Delta$B$_{\rm max}$ = 13\,mmag peak-to-peak) among the 
stars in our sample, and was at first observed for 12 consecutive pulsation cycles in April 2002. Thirty series 
with exposure times of 60--80\,s were acquired and about 4.5\,min were spent for a single measurement of the 
magnetic field. The highest peak in the amplitude spectrum (Fig.~\ref{fig:hd101065}, upper panel) is at a 
frequency of 1.365\,mHz with an amplitude of $39 \pm 12$\,G. This 3.2$\sigma$ signal is at the known photometric 
pulsation frequency, hence is possibly real. The False Alarm Probability of a 3.2-$\sigma$ peak somewhere in the 
amplitude spectrum is high, since many independent frequencies have been searched. But the False Alarm 
Probability at a {\em particular} frequency, namely the known
pulsation frequency, is much lower. 
% GM: the above is quite vague. It should be possible to give
% quantitative estimates of the False Alarm Probability in both
% cases. 

The theoretical considerations of pulsationally-modulated magnetic field variations in case of the star 
HD\,101065 show that the pulsation could give rise to magnetic field variations up to $10^2$\,G if we
assume the radial velocity amplitude to be of the order of 1\,km~s$^{-1}$:
$\left|\delta B_\theta/B\right|
\sim\left|\partial\xi/\partial r\right|\sim P\,\delta
V/(2h\,\pi)\sim0.1$,
hence $\delta B_\theta\approx 100$\,G, in this case,
consistent with this possible detection.
However, we should note that no study of radial velocity variations 
over the pulsation cycle has ever been carried out for HD\,101065.
With this encouraging indication four additional hours with FORS1 were granted by the Director's Discretionary 
Time Committee (Programme No.~270.D-5023) to test our possible
detection. The 72 additional measurements 
were obtained in March 2003;
they are 
presented in Table~\ref{tab:hd101065b}. 
Unfortunately, we did not confirm the result, as can be seen in the
lower panel of Fig.~\ref{fig:hd101065}. The new observations do not rule out the possibility of variation of 
the magnetic field with the known photometric frequency, but they also certainly do not confirm it.

\begin{table}
\caption{
\label{tab:hd101065b}
Mean longitudinal magnetic field for a four hours time series of
HD\,101065 obtained in March 2003.
}
\begin{center}
\begin{tabular}{crr|crr}
\hline
MJD &\multicolumn{1}{c}{$\left<{\cal B}_z\right>$}&\multicolumn{1}{c}{$\sigma{}_z$} 
&MJD&\multicolumn{1}{c}{$\left<{\cal B}_z\right>$} &\multicolumn{1}{c}{$\sigma{}_z$}\\
&\multicolumn{1}{c}{[G]}&\multicolumn{1}{c}{[G]} & &\multicolumn{1}{c}{[G]}&\multicolumn{1}{c}{[G]} \\
\hline
52701.169601 & $-$1031 & 50 &	52701.254376 & $-$1121 & 59 \\
52701.172534 & $-$973 & 49 &	52701.256476 & $-$1081 & 60 \\
52701.175441 & $-$910 & 50 &	52701.258483 & $-$1073 & 62 \\
52701.178343 & $-$987 & 49 &	52701.260479 & $-$1095 & 67 \\
52701.181240 & $-$982 & 49 &	52701.262477 & $-$986 & 66 \\
52701.184140 & $-$964 & 48 &	52701.264478 & $-$1036 & 62 \\
52701.187049 & $-$954 & 48 &	52701.266475 & $-$1019 & 64 \\
52701.189827 & $-$894 & 52 &	52701.268472 & $-$925 & 66 \\
52701.192397 & $-$1072 & 53 &	52701.270472 & $-$1005 & 64 \\
52701.194950 & $-$982 & 53 &	52701.272471 & $-$950 & 65 \\
52701.197503 & $-$889 & 51 &	52701.274468 & $-$1081 & 63 \\
52701.200059 & $-$952 & 51 &	52701.276471 & $-$958 & 68 \\
52701.202527 & $-$835 & 59 &	52701.278473 & $-$1046 & 74 \\
52701.204782 & $-$792 & 59 &	52701.280492 & $-$1062 & 65 \\
52701.207092 & $-$1007 & 56 &	52701.282594 & $-$1115 & 62 \\
52701.209416 & $-$981 & 55 &	52701.284715 & $-$974 & 61 \\
52701.211743 & $-$905 & 54 &	52701.286833 & $-$1041 & 62 \\
52701.214011 & $-$1034 & 57 &	52701.288956 & $-$1002 & 59 \\
52701.216236 & $-$1126 & 57 &	52701.291079 & $-$1080 & 59 \\
52701.218449 & $-$949 & 56 &	52701.293201 & $-$1097 & 62 \\
52701.220655 & $-$887 & 56 &	52701.295327 & $-$1076 & 60 \\
52701.222810 & $-$1135 & 63 &	52701.297456 & $-$960 & 61 \\
52701.224909 & $-$1033 & 58 &	52701.299584 & $-$1077 & 60 \\
52701.227011 & $-$896 & 60 &	52701.301712 & $-$963 & 62 \\
52701.229111 & $-$932 & 58 &	52701.303843 & $-$936 & 59 \\
52701.231212 & $-$1076 & 59 &	52701.305974 & $-$1134 & 60 \\
52701.233315 & $-$987 & 60 &	52701.308101 & $-$972 & 60 \\
52701.235415 & $-$1006 & 56 &	52701.310327 & $-$995 & 58 \\
52701.237517 & $-$1076 & 59 &	52701.312510 & $-$1036 & 70 \\
52701.239618 & $-$1081 & 58 &	52701.314551 & $-$895 & 67 \\
52701.241720 & $-$985 & 60 &	52701.316673 & $-$1061 & 61 \\
52701.243827 & $-$1072 & 61 &	52701.318811 & $-$837 & 63 \\
52701.245936 & $-$1063 & 60 &	52701.320952 & $-$1064 & 61 \\
52701.248043 & $-$1006 & 60 &	52701.323091 & $-$940 & 60 \\
52701.250150 & $-$1024 & 58 &	52701.325230 & $-$1014 & 68 \\
52701.252256 & $-$1090 & 58 &	52701.327429 & $-$991 & 58 \\
\hline
\end{tabular}
\end{center}
\end{table}

{\it \object{HD\,128898}}
($\alpha$\,Cir; HR\,5463) is the brightest roAp star at $V = 3.2$. It pulsates with one dominant 
mode with a  frequency of $\nu = 2.44$\,mHz ($P = 6.8$\,min), and several much lower amplitude frequencies. 
The principal mode is rotationally modulated by a very small amount with the rotation period of 4.463 day 
(Kurtz et al. \cite{KSMT94}).  We obtained 26 series of magnetic measurements with exposure times of 3 to 
4\,s distributed over 9 consecutive pulsation cycles. The highest peak found in the amplitude spectrum 
($\nu = 2.784$\,mHz; $A = 104$\,G) from the measurements obtained in the full wavelength range is not at the 
known photometric pulsation frequency (2.442\,mHz). No significant peaks were found in the amplitude spectra 
from the measurements of the individual Balmer lines. 

\begin{table*}
\caption{
\label{tab:overview}
Rapidly oscillating Ap stars studied for magnetic field pulsations
}
\begin{center}
\begin{tabular}{rlccccccc}
Star & $T_{\rm eff}$ &$\log~g$ & $v\,\sin i$&$P_{\rm rot}$&$\left< {\cal B}_z\right>_{\rm aver}$& 
$\sigma_{z,{\rm aver}}$ & $P_{\rm puls}$ &$\Delta$B$_{\rm max}$\\
 &\multicolumn{1}{c}{[K]} & & [km s$^{-1}$] & [d]&[G]&[G]&[min]&[mmag]\\
\hline 
HD\,83368$^1$ & 7960 & 4.0 & 33.0 & 2.85&$-$847 $\pm$ 108& 56 & 11.6&10\\
HD\,101065$^2$ & 6600& 4.2 & 3.5 & &$-$1014 $\pm$ 72& 57 & 12.1&13\\
HD\,128898$^1$ & 7998& 4.2& 13.0 & 4.48&$-$239 $\pm$ 136& 62 &6.8&5\\
HD\,137949$^1$ & 7030& 3.9 & & $>$ 75\,y?& 2146 $\pm$ 55& 51 & 8.3&3\\
HD\,201601$^1$ & 7620 & 4.0 & & $>$ 70\,y&$-$1072 $\pm$ 173& 55 &12.4&3\\
HD\,217522$^3$ & 6808& 4.0 & 3.0 &&$-$686 $\pm$ 111& 79 & 13.9&4\\
\hline
\end{tabular}
\end{center}

$^1$Hubrig et al. (\cite{HNM00});
$^2$Cowley et al. (\cite{CR00});
$^3$Hubrig et al. (\cite{HC02}).

\end{table*}

{\it \object{HD\,137949}}
(33 Lib) pulsates in one mode with a  frequency of $\nu = 2.01$\,mHz ($P = 8.3$\,min) 
(Kurtz \cite{K91}). This star exhibits the strongest longitudinal magnetic field among the stars of our sample. 
We obtained 30 series of magnetic measurements with exposure times of 30\,s over 10 pulsation cycles. No peak 
at the known photometric frequency (2.014\,mHz) was found in the amplitude spectra of the magnetic field 
measurements obtained in the full wavelength range; the highest noise peaks had amplitudes of only 30\,G. No 
significant peaks were found in the amplitude spectra from the measurements of the individual Balmer lines. 

{\it \object{HD\,201601}}
($\gamma$\,Equ; HR\,8097) is the second brightest roAp star at $V = 4.7$. It pulsates with four 
frequencies with periods near 12.3\,min (Martinez et al. \cite{M96}). Magnetic measurement data were taken 
sporadically because of clouds passing during the observations. We obtained in all only 18 series with exposure 
times of 2 to 5\,s. No signal was detected, and, as a consequence of the sparseness of the data, the highest 
noise peaks in the amplitude spectra are at 170\,G –- far higher than in our data sets for other stars. No 
significant peaks were found in the amplitude spectra from the measurements of the individual Balmer lines.

{\it \object{HD\,217522}} (CPD $-$45\degr10378)
was originally discovered to pulsate in a single mode with a period of 13.7-min (Kurtz \cite{K83}). 
Further observations by Kreidl et al. (\cite{KKB91}) found another pulsation mode with a period of 8.3\,min 
that was not present in the discovery data set, indicating transient modes, or strong amplitude modulation. 
Magnetic measurements were taken over 24 consecutive cycles (of the original mode) with 91 series with the 
exposure time of 30\,s. We found in the amplitude spectrum obtained from the measurements in the full 
wavelength range a highest peak with an amplitude of 36\,G at 1.673\,mHz corresponding to a period of 9.97\,min. 
This peak is not significant and does not match any known photometric pulsation period. No significant peaks 
were found in the amplitude spectra from the measurements of the individual Balmer lines.

\section{Discussion}\label{sec4}

The basic data of our sample of roAp stars are listed, star by star, in Table~\ref{tab:overview}. Successive 
columns give the HD number, the effective temperature, the gravity,
the value of $v\,\sin i$ and of the rotation 
period, if known. The mean value, for every star, of all the longitudinal field
measurements of Tables~\ref{tab:hd83368} to \ref{tab:hd101065b},
$\left< {\cal B}_z\right>_{\rm aver}$, and
the standard deviation of the individual measurements about this mean,
appears in Col.~6. The mean of all the standard deviations of the
individual measurements, $\sigma_{z,{\rm aver}}$ from the same tables,
is presented in Col.~7. 
In cols. 8 and 9 we list the photometric pulsation period $P_{\rm puls}$ and the photometric 
pulsation amplitude $\Delta B_{\rm max}$. The indexes 1--3 in the HD 
number column refer to the literature sources of the atmospheric parameters.

With UT3 of the VLT and FORS\,1, we measured the mean longitudinal field variation over the pulsation cycle in 
six roAp stars to begin the study of how the magnetic field and pulsation interact. Only the star HD\,101065, 
which has one of the highest photometric pulsation amplitudes, showed a potential signal at the known 
photometric pulsation frequency at the 3$\sigma$ level. We found a signal for magnetic variability with a 
frequency of 1.365\,mHz and an amplitude of 39$\pm$12\,G in this star. First theoretical considerations led to 
the result that in case of the star HD\,101065 the pulsation could give rise to magnetic field variations
 $\left|\delta B_\theta \over B\right|
\sim0.1 \rightarrow \left|\delta B_\theta\right|\approx 100$\,G, 
% \sim\left|{\partial\xi\over\partial r}\right|\sim{P\delta V\over2h\pi}
% GM: I think its useless to repeat one more time the whole sequence
% of conversions from velocity variations to field variations (which
% already appears at least twice elsewhere in the paper). 
consistent with our possible detection. However, our second attempt to measure magnetic 
variability in this star during 4 hours failed to obtain a positive detection again. Given the noise level, the 
amplitude spectrum of the new observations does not rule out the possibility of variations at the level found 
from the measurements in April, but it certainly does not confirm them. 

Our study has come close to the limits of what is currently feasible for the measurements of magnetic fields in 
roAp stars with FORS\,1 at the VLT. Although the low resolution spectropolarimetry of hydrogen Balmer lines 
obtained with FORS1 represents a powerful diagnostic tool for detection of stellar magnetic fields, the 
accuracy of spectropolarimetric measurements with FORS\,1 has not yet been quite high enough to detect magnetic 
pulsation variations of the order of a few tens of G that we expect to be present. Given our possible marginal 
detection with a few hours of observations of HD\,101065, an 8\,h run over about 40 pulsation cycles should 
increase our signal-to-noise ratio by nearly a factor of two and show a clear result.

{\change Because of the low resolution of the FORS\,1 spectra, no magnetic field measurements can be 
carried out using exclusively doubly-ionized rare-earth lines which show the largest pulsational radial 
velocity amplitudes in roAp stars. In general, pulsational amplitudes found in the hydrogen lines are lower 
than for doubly-ionized rare-earth lines
implying a lower level of the expected magnetic field variations.
The advantage of using high-resolution spectropolarimetry has been demonstrated 
in the very recently published study of magnetic field variations over  
the pulsation period in the roAp star $\gamma$\,Equ by Leone \& Kurtz (\cite {LK03}). A series
of spectra has been obtained on the 3.55\,m 
Telescopio Nazionale Galileo (TNG) with the high resolution spectrograph SARG equipped with a polarimeter.
Magnetic field measurements of four strong lines of \ion{Nd}{iii}
showed a variability over the pulsation cycle with an amplitude in the range 112--240\,G.

As mentioned in the introduction, the primary purpose of our work was to obtain observational
information on magnetic field variations and its very interest was to provide constraints for further 
development of theoretical models.
In this respect, it is noteworthy that adopting the measured radial velocity amplitude of 
the \ion{Nd}{iii} lines of the
order of 400\,m\,s$^{-1}$, the expected magnitude of magnetic field variations for $\gamma$\,Equ according to 
the theoretical estimate of Sect.~1 is at least two times lower than the magnitude 
observed by Leone \& Kurtz.} 

To summarize,
from the observational point of view there is a need to observe more roAp stars in order to understand how the 
pulsation pattern interacts with the magnetic field on the time-scale of the pulsations. The approach should 
consist in achieving better accuracy of magnetic field measurements taking advantage of the large light 
collecting power of 4\,m to 8\,m class telescopes and high-resolution spectropolarimeters like 
SARG on the 3.55\,m TNG telescope or ESPaDOnS which 
will be installed onto the Canada--France--Hawaii telescope this year. In the absence of spectropolarimetric 
instruments on large telescopes, this study can be complemented by high spectral resolution, high time resolution 
spectra of roAp stars with magnetically split lines obtained in unpolarized light at 4\,m to 8\,m class 
telescopes. Such observations, which are sensitive to the
mean magnetic field modulus rather than the mean longitudinal magnetic
field considered here, can potentially detect lower
amplitude variations of this field moment (uncertainties of individual
measurements based on a single line can be as low as $\sim25$~G).
Observations of this type using the VLT have already been carried out
for one roAp star (Mathys, Kurtz \& Elkin, in preparation).

\end{document}